\documentclass[%
 reprint,%
 amssymb, amsmath,%
 aip,jap,%
]{revtex4-1}
\usepackage{graphicx}% Include figure files
\usepackage{dcolumn}% Align table columns on decimal point
\usepackage{bm}

\draft 

\begin{document}

\preprint{AIP/jap}

\title{On the Geometry of Carbon Nanostructures Formed at
Reaction of Organic Compounds at High Pressure and Temperature}% Force line breaks with \\
%\thanks{Footnote to title of article.}

\author{Nataliya P. Satonkina}
 \protect\thanks{Electronic mail: snp@hydro.nsc.ru.}
\author{Dmitry A. Medvedev}
\affiliation{Lavrentyev Institute of Hydrodynamics, 630090 Novosibirsk, 630090 Russia}\affiliation{ Novosibirsk State University, Novosibirsk, 630090 Russia}

%\textit{$^{a,b}$}

\date{\today}

\newcommand\revtex{REV\TeX}

%\preprint{}

% ail, \thanks, \homepage, \altaffiliation all apply to the current author.
% actual e-mail address or url should go in the {}'s for ail and \homepage.


%ail[]{Your e-mail address}
%\thanks{}

\begin{abstract}
Based on the analysis of the data on the behavior of electric conductivity at the detonation of condensed high explosives (HEs) with the composition CHNO and the carbon mass
fraction higher than 0.1, the conclusion was made of the presence of long carbon nanostructures. These structures penetrate all the space of reacting HE. The structures are formed already in the chemical peak region, and they evolve along the detonation wave.
\end{abstract}

\maketitle 

\section{Introduction}
Condensed carbon of different modifications is released at the detonation of high explosives with the composition C$_a$H$_b$N$_c$O$_d$ and negative oxygen balance. Ultrafine diamonds were found in conserved detonation products (DP).\cite{bib:lyamkin88,bib:philips88} Despite the serious age of this discovery, there still remain several questions which provoke intense discussion. There is no common opinion about the shape of carbon particles at the carbon condensation during the detonation, about the condensation proceeding, and about the moment of formation of single particles.

In the literature, there are different viewpoints on the geometry of carbon inclusions. Results
obtained by different investigation methods lead to different conclusions. In the work,\cite{ers1991} carbon condensation to single particles was simulated with further formation of
fractal structures from these particles. The mechanism restricting the growth of single particles
was proposed. In the work,\cite{bib:satonkina14} extended structures were obtained in a numerical
experiment on the condensation of carbon. When the mass fraction of carbon is higher than 0.1, the
stage of single particles is absent, and the condensation produces directly the extended
structures. The electric conductivity of these structures was calculated which agrees well with the
experimental data.

The viewpoint of the existence of single particles is shared by many authors (see for example.\cite{zubkov,sorin,sorin2001,sorin2017,gorshkov,bib:danilenko051}) In the work,\cite{zubkov} a dynamic model of
carbon condensation at the detonation of trinitrotoluene (TNT) was proposed based on the results of
the electric conductivity investigation. The author of work\cite{sorin} considered the energy release
at the condensation and obtained the growth of single carbon particles during several microseconds.
In the work\cite{gorshkov} on the behavior of the electric conductivity at the detonation of
triaminotrinirobenzene (TATB), the growth of carbon particles in the Taylor wave was considered
supposing the thermal electric conductivity mechanism.

There are several works where the data obtained using the synchrotron radiation are interpreted as
an existence and growth of single carbon particles to the size of several tens of micrometers
during several microseconds (for example, work\cite{sax2012}). On the contrary, it was found in the
work\cite{hns} using the same method that at the detonation of hexanitrostilbene
C$_{14}$H$_6$N$_6$O$_{12}$), single particles appeared faster than in 0.5 microseconds, and the
particle size of 2.7 nm remained constant during the whole measurement time (3 microseconds).

Thus, different investigation methods reveal both the extended structures which are hard to detect
by the small angle x-ray scattering, and single particles found in conserved detonation products.
The model of carbon condensation at the detonation is absent.

High pressure and temperature at the detonation of condensed HEs restrict the
circle of experimental investigation methods, and the interpretation of the data obtained is
ambiguous. At the present stage, comparison of the available experimental data is the most
effective method. Under the assumption of a connection of electric properties and the presence of carbon,\cite{bib:satonkinajap,sat162} the data on electric conductivity is the ultimate source of
information.

TNT has the highest amount of carbon among all HEs, the carbon mass fraction in molecule is 0.37,
and the fraction of free carbon in the CJ point is 0.27.\cite{bib:tanaka} Further we will discuss
that the electric conductivity in TNT can be satisfactorily explained by carbon nets.

For other HEs such as TATB, RDX, octogen (HMX), and pentaerythritol tetranitrate (PETN), the
relation between carbon and conductivity was not considered until recently due to relatively
small amount of condensed carbon. It was proposed in works\cite{bib:satonkinajap,sat162} to explain
the maximum value of conductivity by the formation of carbon structures in the chemical peak.
Hence, it was supposed that carbon is fully condensed before the CJ point, and free carbon in
detonation products is the remnant after the end of chemical reaction. In the framework of this
model, the uniform dependence was obtained for the maximum conductivity and the conductivity in
the CJ point for five HEs.

Thus, the condensation of carbon is inseparably connected to the electric conductivity. In this
paper, we discuss the most realistic hypotheses of conductivity at the detonation of condensed HEs.
It is shown that only the contact mechanism can explain the experimental data. HEe with the mass
fraction of carbon higher than 0.1 are considered (TNT, RDX, TATB).

\section{Experimental data on the behavior of electric conductivity at the HE
detonation}

\begin{figure}[h!]
 \includegraphics[width=85mm]{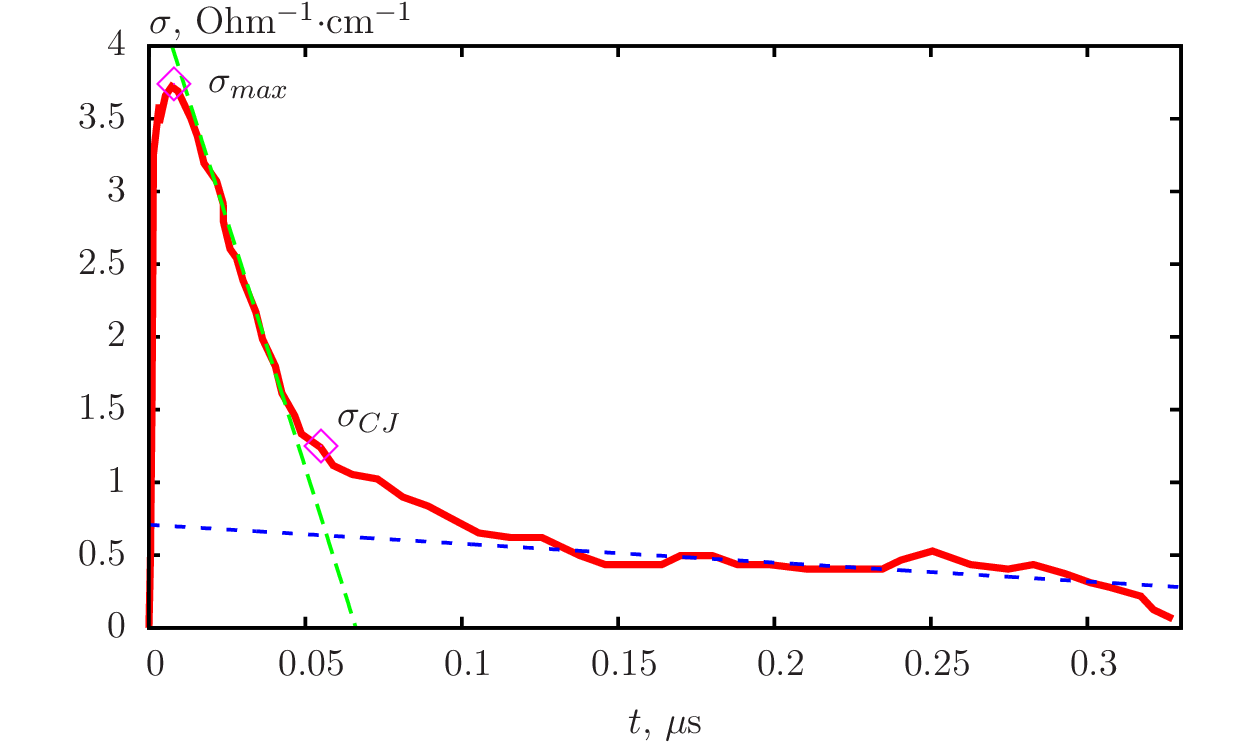}%
 \caption{Graph of conductivity at the detonation of RDX.\label{g97}}%
\end{figure}

The detonation wave consists of the shock front, the adjacent chemical peak, and the Taylor
rarefaction wave which is separated from the chemical peak by the Chapman -- Jouguet (CJ) plane.

Typical electric conductivity profile is shown in Fig. 1. There is a fast growth to the maximum
value, then a decrease with a gradient dependent on the explosive type, and after the inflection
point, a slow variation in the Taylor wave with small value. Maximum conductivity $\sigma_{max}$
corresponds to a region inside the chemical peak. The inflection point $\sigma_{CJ}$ at the
$\sigma(t)$ graph is related to the CJ point.\cite{bib:satonkinajap,sat162}

Despite the absence of a complete conductivity theory, it is possible using the experimental data
to exclude conductivity mechanisms which are not decisive for the electric properties of the
detonation wave, and to mace some conclusions about the geometry of carbon inclusions.

In the table,\ref{temptab} the data for HEs based on TATB, TNT, and RDX are presented. The
notation is following: $\rho$ is the initial denstiy, $T_{CJ}$, $P_{CJ}$ and $r_{CJ}$ are
temperature, pressure and the mass fraction of condensed carbon in the CJ point, $r_{c}$ is the
carbon fraction in molecule, $\sigma_{CJ}$ and $\sigma_{max}$ is the conductivity in the CJ point
and the maximum conductivity, correspondingly, $d_{cr}$ is the critical diameter. Temperature,
pressure and mass fraction of condensed carbon in the CJ point were obtained numerically in work.\cite{bib:tanaka}

ll the HEs chosen are well known and intensely investigated. They have drastically different
characteristics such as sensitivity, temperature and pressure in the detonation wave. The
parameters of the detonation wave for HMX are close to those of RDX and PETN.

Presently, the most popular viewpoint is that in HEs with a slightly negative oxygen balance (PETN,
RDX, HMX), the conductivity in the chemical peak could be related to the chemoionization analogous
to the process of combustion \cite{bib:ershov07} where free electrons with high mobility appear as
a result of chemical reaction. In the work,\cite{pe} besides the chemoionization, the ionization
of intermediate detonation products (DP), the dissociation of PD, and the thermal emission from
carbon particles are listed as a source of charge carriers/ The conductivity both in the chemical
peak and in the Taylor wave are explained by the thermal emission. The existence of a correlation
between the conductivity $\sigma$ at the CJ point and the amount of condensed carbon was shown in works.\cite{bib:hayes65,bib:gilev2001} There exist also works where the value of $\sigma_{CJ}$ is
explained by the ionic conductivity \cite{bib:ershov09} and by the thermal ionization.\cite{gorshkov}

Let us consider different conductivity mechanisms from the standpoint of the experimental data presented.

\begin{center}
\begin{table*}
\caption{ }
\label{temptab}
  \begin{tabular}{c c c c c c c c c c}\hline  \hline
 N  &  HE & $\rho$,& $T_{CJ}$, K& $P_{CJ}$, kbar&  $r_{CJ}$  &$\sigma_{CJ}$,& $r_c$ &$\sigma_{max}$,&$d_{cr}$, mm\\
    &     &g/cm$^3$&    &              & &Ohm$^{-1}$cm$^{-1}$ &  &Ohm$^{-1}$cm$^{-1}$&  \\  \hline
 1  & TATB & 1.8    & 2762& 267 &  0.208     &10.0&     0.279&19.1  & 6.35 \cite{LLNL}  \\
    &      &        &           &            &    &          &      &    &  ($\rho\approx 1.7$)   \\ %\hline
  2  & TNT &  1.0    & 3398& 71  &  0.128     &8.9 &     0.370 &15.0 & 3$\div$5 \cite{pe}  \\
    &      &         &     &     &           &     &           &     & ($\rho\approx 1.51$)  \\ %\hline
 3  & TNT &  1.6    & 3434& 191 &  0.260     &26.8&     0.370 & $\sim$100  &  16 \cite{dremin}  \\
    &      &         &     &     &           &     &           &     & ($\rho=1.62$)  \\ %\hline
 4  & RDX &  1.2    & 3964& 148 &  0.030     &0.4 &     0.162&1.8   & $2 \div 4$ \cite{pe}  \\
     &      &         &     &     &           &     &           &     &  ($\rho\approx 1$)  \\ %\hline
5  & RDX &  1.6    & 3675& 252 &  0.066     &1.25&     0.162&4.2   &   \\
 \hline  \hline
\end{tabular}
\end{table*}
\end{center}

\section{Discussion}

\subsection{Chemoionization}

It was established by the comparison of experimental data on the duration of the reaction zone and
the region of high conductivity that the maximum value of conductivity $\sigma_{max}$ is reached
inside the reaction zone.\cite{bib:ershov07,bib:ershov09,arxiv16} Chemoionization is thought to be
the most probable cause of conductivity in the chemical peak region. The concentration of charge
carriers $n$ due to chemoionization is by definition connected with the number of reacted
molecules. For the densities lower than the critical one, the speed of decomposition is maximal in
the shock front and then monotonically decreases.\cite{bib:loboiko} Following the argument of
chemoionization, conductivity should be the higher the faster the initial substance decomposes,
that is the maximum of conductivity $\sigma_{max}$ should be close to the detonation front.
According to our data for HEs with well defined conductivity peak, $\sigma_{max}$ is reached at
about the half of duration of the zone of increased conductivity,\cite{bib:satonkina14,bib:ershov07,bib:ershov09,bib:ershov04,bib:ershov10,bib:safonov} i.e., far
from the front which contradicts to the relation of the conductivity $\sigma$ and the intensity of reactions.

On the other hand, the number of elementary chemical reactions per unit time increases with the
decrease of the duration of the reaction zone, and, according to Hariton's principle, the duration
of the reaction zone is proportional to the critical diameter. Overall speed of chemical
decomposition in RDX is close to that in TNT (see Table q). The conductivity in TNT is however
higher by an order of magnitude $\sigma_{max TNT 2}\approx 8\sigma_{max RDX 4}$. For TATB based HE,
the conductivity has an intermediate value $\sigma_{max TNT 2}
>\sigma_{max TATB}>\sigma_{max RDX 4}$ whereas the critical diameter is larger. Thus, the data of
Table one show the absence of the dependence of maximum conductivity on the critical diameter.

The time of existence of free electrons in the chemical peak can play a significant role. If
electrons produced in chemoionization are immediately captured by atoms forming ions, they have no
time to make a contribution into conductivity, and the conductivity would be an ionic one. The
mobility of ions is however insufficient to produce experimentally observed conductivity.\cite{bib:ershov75} In the case of a large time of existence of free electrons, the concentration
of charge carriers and, hence, the conductivity would increase in the course of chemical reaction.
In the detonation wave, pressure and density reach maximum values near the front and then decrease
monotonically.\cite{pe,bib:loboiko,zeldovich1940,fedorov} When pressure and density decrease, the
intensity of recombination decreases,\cite{zeldraiz} the mobility of electrons increases, and the
concentration of reacting component decreases. From this, one would expect that conductivity would
at least not fall at the decrease of pressure and density which is not observed.

Hence, we can conclude that chemoionization does not explain the experimental data. There is no relation
between the speed of chemical decomposition and the maximum value $\sigma_{max}$. The explanation of highly
nonuniform conductivity distribution is also absent.

In the works,\cite{bib:satonkinajap,sat162} the correlation between the maximum value of conductivity
$\sigma_{max}$ and the carbon content in a molecule $r_c$ was demonstrated based on experimental data for five HEs.
The increase of the maximum value $\sigma_{max}$ with the increase of the HE density with the same grain size
was observed. We suppose that this fact is related to the increase of the density of carbon, and it is a crucial
factor for the conductivity.

In our opinion, the conductivity inside the chemical peak region is provided by carbon
nanostructures which are produced at the destruction of molecules displacing other atoms into the
space between structures.\cite{bib:breusov,pccp15} The data of Table \ref{temptab} show the
dependence of the maximum conductivity $\sigma_{max}$ on the mass fraction of carbon $r_c$, and the
absence of relation with the critical diameter and the chemoionization.

\subsection{ Relation between electric conductivity and temperature}

In many works, the conductivity at the detonation is related to the high temperature. Temperature is however
the most badly defined among all detonation parameters. Results of both calculations and experiments give
sometimes even qualitatively different behavior of $T(\rho)$. Nevertheless, it is commonly accepted that the
temperature at the detonation of TNT is lower than one at the detonation of RDX, HMX, PETN.

The relation between temperature and conductivity can be provided by thermal ionization and thermal
emission. Thermal ionization is the appearance of charge carriers (free electrons and ions) due to the
high temperature. Thermal emission is the emission of free electrons from the surface of carbon particles.
The temperature dependence of the ionic conductivity is qualitatively the same as of the electronic one.
In the case of the same concentration of ions and electrons, the electronic conductivity would dominate
due to higher mobility of charge carriers. Hence, we consider here electrons as charge carriers.

The temperature dependence of the conductivity for both the thermal ionization and thermal emission is
exponential $\sigma\sim  n_e \sim \exp(-\frac{E}{kT})$ with the activation energy of order of several eV.\cite{kikoin} The multiplier dependent on pressure, concentration of carriers, etc. changes for different
only slightly.

\begin{figure}
 \includegraphics[width=85mm]{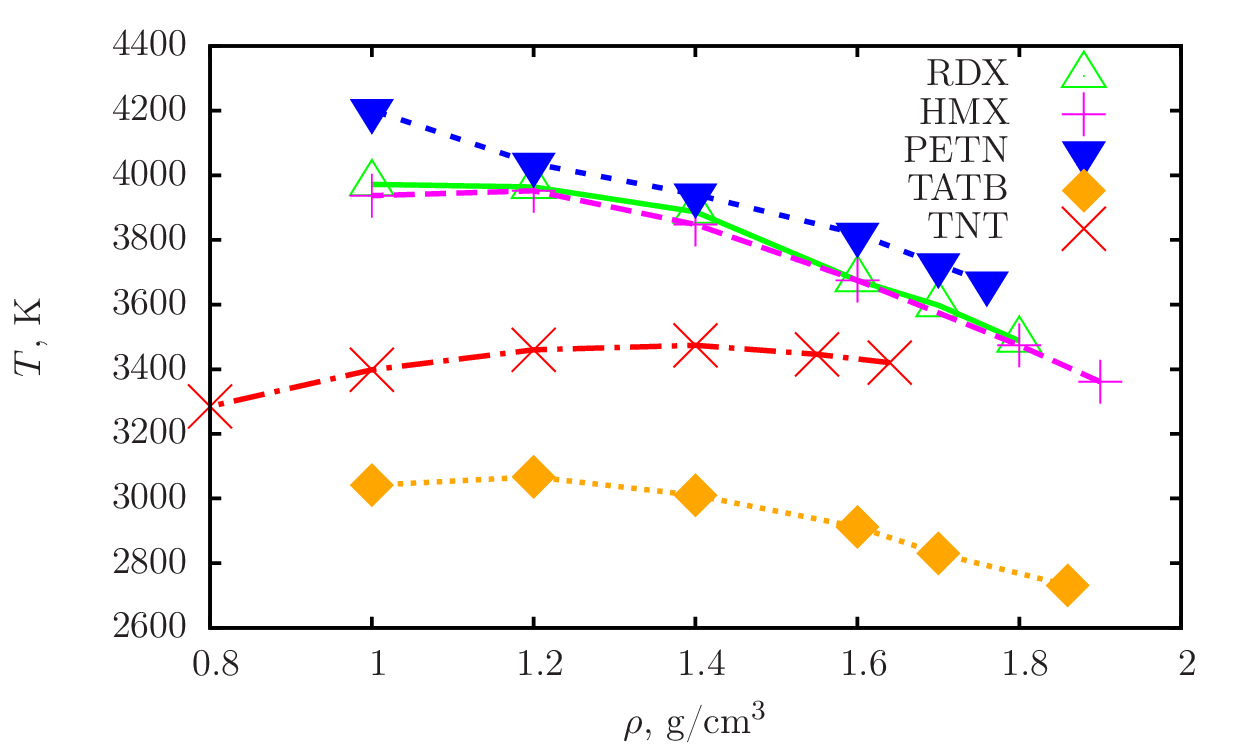}%
 \caption{Temperature at the CJ point, data from work\cite{bib:tanaka} \label{temp}}%
\end{figure}

It was obtained experimentally that for three of listed in Table 1 HEs the conductivity increases
with the increase of density.\cite{bib:satonkina14,bib:ershov07,bib:ershov09,bib:ershov04,bib:ershov10,bib:safonov} The
behavior of temperature is more complex. Figure \ref{temp} shows the temperature in the CJ point
for different density of HE.\cite{bib:tanaka} For RDX, HMX and PETN, temperature decreases
slightly with the increase of density. For TNT, the dependence $T(\rho)$ is non-monotonous given
the same values of $T$ at different densities (Table 1, TNT$_2$ and TNT$_3$). Since the mobility of
electrons increases and recombination processes slows down at the decrease of pressure and density\cite{zeldraiz} (in our case for TNT$_2$), it could be expected that $\sigma_2$ should be larger
than $\sigma_3$. Experiments give however that $\sigma_{CJ2}/\sigma_{CJ3}=1/3$,
$\sigma_{max2}/\sigma_{max3}\approx1/6$, the values for smaller density are lower. This contradicts
the assumption of the thermal nature of conductivity

The increase of conductivity with the increase of density can not be explained by temperature which
is lower for RDX at higher density. This also supports the non-thermal origin of conductivity.

Besides, the temperature for RDX is higher than the one for TNT by $\sim 450$ K at significantly
lower values of $\sigma$. The temperature for TATB based HE, the temperature is lower than the one
for TNT by $\sim 500$ K which does not produce a strong exponential dependence in experimental data
(table \ref{temptab}).

It was noticed earlier in work\cite{bib:ershov75} that the estimate of the degree of thermal ionization by the Saha
formula gives the value of $\sim10^{-6}$ which does not explain the experimental data on conductivity.

The thermal emission was investigated in detail in the work.\cite{bib:ershov75} Following
phenomena influencing the thermal emission were considered: the growth of carbon particles with
time, the increase of the work function with increasing size of a carbon particle, the decrease of
the work function due to interaction with dense ambient products. The interaction of electron with
surrounding positively charged carbon particles was also taken into account. It was obtained that
the concentration of free electrons in the Taylor wave in TNT can be as high as $10^{19}$~cm$^{-3}$
due to the thermal emission. This value is ten times lower than the one estimated from experimental
value of conductivity $2-4$~Ohm$^{-1}$cm$^{-1}$. It was noted that the value obtained from
experiments is close to the estimated based on thermal emission. It was impossible to make a choice
between the thermal emission and the contact mechanism. In the works,\cite{bib:hayes65,bib:ershov09,bib:gilev} maximum conductivity at the detonation of TNT was
$100-250$~Ohm$^{-1}$cm$^{-1}$ which is two orders of magnitude higher than the value obtained in work.\cite{bib:ershov75} The values of conductivity in the work \cite{bib:ershov75} were underestimated
due to imperfect experimental method. Thus, necessary concentration of electrons should be
$10^{22}$~cm$^{-3}$, and the thermal emission can not be a satisfactory explanation. Such a high concentration of electrons is not present at the detonation.

Hence, thermal ionization and thermal emission are certainly present in the detonation wave but
they are not the determining factor for the conductivity. Table \ref{temptab} shows that the
increase of density and carbon fraction influence the conductivity more strongly than temperature
value.

\subsection{Contact mechanism of electric conductivity}

In the works of Gilev,\cite{bib:gilev,bib:gilevdis} the maximum value of conductivity at the
detonation of TNT was estimated from a percolation model. It was obtained that even the total
amount of carbon is not sufficient to explain observed values, and highly conductive elongated
structures need to exist already in the chemical peak region.

Let us estimate the conductivity of a cube with a content of carbon equal to that in TNT of maximum
density, i.e., $m_C=\rho_{max}*r_c=1.6*0.37=0.6$ g. The conductivity of carbon varies in a broad
range, from dielectric diamond to almost metallic one for highly-oriented graphite. Conductivity of
graphite depends on the modification of the crystal lattice and on the crystallographic
orientation, and it varies from $\sim 244$ to 1250~Ohm$^{-1}$cm$^{-1}$.\cite{spr1,spr2} The
conductivity of highly-oriented graphite can be as large as 20000~Ohm$^{-1}$cm$^{-1}$.\cite{korobenko99} The liquid state of carbon is possible at the detonation
\cite{bib:danilenko051} which has conductivity $\sim 1000$~Ohm$^{-1}$cm$^{-1}$.\cite{korobenko99}
A carbon rod with a mass of 0.6~g and a length $l=1$~cm with a conductivity of carbon
$\sigma_C\approx 1000$~Ohm$^{-1}$cm$^{-1}$\cite{kikoin} has effective conductivity $\sigma\approx
240$~Ohm$^{-1}$cm$^{-1}$ which is close to the experimental data. This value describes the process
better than the estimates based on the thermal emission. These calculations support the hypothesis
on a contact origin of conductivity at the detonation of condensed HEs with electric current along
penetrating carbon nanostructures both in the chemical peak and in the Taylor wave.

Molecular dynamics simulations show that the details on interparticle interaction do not influence
significantly the time of coagulation. Aggregation of carbon atoms is determined by diffusion and
proceeds in picoseconds.\cite{pccp15,bib:satonkina141} At thermodynamic parameters characteristic
for the detonation, the state of carbon corresponds to the condensed phase,\cite{sorin2001,bib:danilenko051}
in contrast to other substances (N, N$_2$, H$_2$O, H, O, CO, CO$_2$, etc.). Probability of
aggregation at a collision of two carbon atoms is close to 100\%, whereas reactions of gaseous
substances under the detonation conditions require of order of 1000 collisions for one elementary
reaction.\cite{zeldovich1940}

In the works \cite{bib:satonkina14,bib:satonkina141}, it was obtained that at the carbon mass fraction higher than
0.1, the coagulation of carbon proceeds immediately to branched nanostructures and not to single particles with
further formation of fractals as supposed in work.\cite{ers1991} The already present in an HE molecule united carbon atoms
will enhance this process (for example, benzene ring in TNT or cross-like structure in PETN).

 Following works confirm the fast aggregation of carbon atoms. Authors of works\cite{bib:anisichkin07,anisichkin16,bib:breusov,pccp15} assert that the aggregation of carbon
occurs already in the reaction zone. V.F. Anisichkin obtained in work \cite{bib:anisichkin07} that
carbon components of molecules of fine-grained mixture HEs completely mix before the oxidation,
i.e., inside the reaction zone, and the oxidation occurs later. O.N. Breusov demonstrated from
energy considerations\cite{bib:breusov} that the formation of carbon clusters is related to the
partial breakup of chemical bonds in molecules and to the formation and growth of carbon skeleton.
In the work,\cite{pccp15} the clustering of carbon at the heating of molecules of TATB, HMX and
PETN was simulated by the molecular dynamics. Carbon nanostructures in the reaction zone were
obtained. Filamentary structures found in conserved detonation products
\cite{vol1,vol2} can be the remnants of fractal structures formed at condensation in the region of
chemical peak. Hence, we can suppose that long carbon filaments of the diameter of 10--40 nm exist
in the detonation wave.

Carbon is able to condense in elongated structures with fractional dimension. Filamentary
structures were found in conserved detonation products.\cite{vol1,vol2} Foam-like structures were
obtained in inert atmosphere at low pressure.\cite{bib:pena,bib:pena1} In the work,\cite{bib:pinaev} formation of nets with a dimension 2.2 was observed, the formation of carbon
clusters at combustion and detonation of gas mixtures was investigated, and clusters of different
modifications we obtained, from fullerene-like to long branched structures made of carbon
particles.

Under a certain fraction of carbon, the formation of penetrating structures becomes impossible due
to the lack of sufficient amount of conductive substance. In the works,\cite{bib:satonkina14,bib:satonkina141} the threshold volume fraction of carbon at which the
formation of connected nets in DP is possible was obtained in numerical experimen to be about 0.07.
Carbon however influence the conductivity even at the fraction lower than 0.07.\cite{bib:satonkinajap,sat162} In such case, the existence of another carbon-related conductivity
mechanism is possible, for example, highly-conductive carbon inclusions in a medium with low
conductivity. When the carbon fraction and the contact conductivity decrease, the role of the ionic
mechanism increases.

In an extreme case of the detonation of gas mixtures, a great progress in understanding the
electric properties was made. In the work,\cite{gas} the conductivity model due to thermal
ionization was considered, the improved by considering quantum-mechanical effects model of
\cite{bib:ershov75} was proposed, and a good agreement with experimental data was obtained. The
model used is based on the presence of free electrons in detonation products. The formula applied
was derived under assumption of rare collisions which is valid for gas detonation but not
applicable for condensed HEs with three orders of magnitude higher density. Despite the external
similarity of the detonation process in gases and in condensed HEs, the nature of electric
properties of these media is different, and the values of conductivity differ by three orders of magnitude.

Let's summarize. Under extreme conditions, carbon in organic compounds tends to aggregate in elongated structures. Atoms of nitrogen, oxygen and hydrogen chemically bound with carbon are present at the surface of such structures.\cite{aggreg} 

With the present state of investigation technology, it is impossible to observe
directly elongated carbon structures during the detonation process directly due to the very short
duration of the process and the aggressive media. The later factor hinders the revealing of such
structures since they get thinned and broken in a chemically active media. Only conserved
detonation products are available for investigation which provide only an overall information.
Thus, the analysis of the electric properties measured at the detonation process seems to be a very
promising and so far an ultimate investigation method.

The results of the present work are useful not only for the study of detonation
and kinetics of HEs. The investigation of the behavior of media under extreme conditions
characteristic for the explosion is rather restricted in methods, and the explosion becomes an
ultimate approach. The results are useful for the fast developing interdisciplinary science, the
abiogenesis\cite{abiogenesis} (formation of organic compounds at the early stage of the origin of life at young
Earth, abiogenic formation of organic molecules characteristic for living organizms). Under high
pressure and temperature, the formation of carbon chains and hydrocarbons occurs. The skeleton of all organic molecules is made of carbon-carbon bonds.
Since carbon is one of the indispensable components of the living matter, the condition of
explosion can model the conditions of meteoritic impacts, planetary depths, volcanos. Hence, the work was useful for astrobiology\cite{astrobiology} and geochemistry.

\section{Conclusion}

Experimental data on the electric conductivity are used as the diagnostic tool and
the indicator of the state of organic compounds under high pressure and temperature. The
consideration of the most realistic hypotheses of conductivity shows that only the contact
mechanism can explain high conductivity values obtained in experiments. This leads to the
conclusion of existence of spatial carbon structures which form a connected nets. Such nets
penetrate all the volume of reacting media and enable the flow of electric current when voltage is
applied. Chemical reactions lead to the thinning and disruption of the structures producing
individual carbon formations which are found in conserved detonation products.

 %You need to replace "rsc" on this line with the name of your .bib file
%\bibliographystyle{rsc}

\begin{acknowledgments}
The work was supported by RFBR, Grant N 15-03-01039.
\end{acknowledgments}

% If in two-column mode, this environment will change to single-column format so that long equations can be displayed. 
%\begin{widetext}
%$$\mbox{put long equation here}$$
%\end{widetext}

% Figures should be put into the text as floats. 
% Use the graphics or graphicx packages (distributed with LaTeX2e).
% Put the label that you will use with \ref{} command in the braces of the \label{} command.
%
% \begin{figure}
% \end{figure}

% \begin{table}
% \end{tabular}
% \end{table}


\end{document}